\documentclass[a4paper,12pt,onecolumn]{IEEEtran} 
\usepackage{setspace}

\usepackage{graphicx}
\usepackage[ruled,vlined]{algorithm2e}
\usepackage{amsmath,amssymb}
\usepackage{color}

\newtheorem{definition}{Definition}{\bfseries}{\rm}
\newtheorem{proof}{Proof}{\bfseries}{\rm}
\newtheorem{theorem}{Theorem}{\bfseries}{\rm}
\newtheorem{lemma}[theorem]{Lemma}{\bfseries}{\itshape}
\newtheorem{example}{Example}{\bfseries}{\rm}
\newtheorem{proposition}{Proposition}{\bfseries}{\itshape}
\newtheorem{remark}{Remark}{\itshape}{\rm}

\newcommand{\newadd}[1]{{\color{black}{#1}}}

\newcommand{\comment}[1]{}

\renewcommand{\emptyset}{\varnothing}

\newcommand{\mc}[1]{\mathcal{#1}}

\newcommand{\mb}{\mathbb}

\newcommand{\mP}{\mathcal{P}}

\newcommand{\mD}{\mathcal{D}}

\newcommand{\mM}{\mathcal{M}}

\newcommand{\mF}{\mathcal{F}}

 \newcommand{\mv}[1]{\mathrel{\stackrel{#1}{\rightarrow}}}

\newcommand{\Act}{\textit{Act}}

\newcommand{\Nats}{\mathbb{N}}
\newcommand{\Reals}{\mathbb{R}}

\def\topbotatom#1{\hbox{\hbox to 0pt{$#1\bot$\hss}$#1\top$}}

\makeatletter
\newcommand{\be}{%
\begin{group}
\eqnarray%
\@ifstar{\nonumber}{}%
}

\makeatother

\newcommand{\Power}{\mathcal{P}}
\newcommand{\supp}{\mathsf{Supp}}

\begin{document}
\title{Polynomial-time Algorithms for Computing Distances of Fuzzy Transition Systems}
\author{Taolue Chen, Tingting Han, and Yongzhi Cao, {\it Senior Member, IEEE}%
\thanks{T. Chen is partially supported by EPSRC grant (EP/P00430X/1). Y. Cao was supported by National Natural Science
Foundation of China (Grant No. 61370053).}%
\thanks{T.~Chen is with the Department of Computer Science, Middlesex University London, The United Kingdom (e-mail: t.chen@mdx.ac.uk).}
\thanks{T.~Han is with Department of Computer Science and Information Systems, Birkbeck, University of London, The United Kingdom (e-mail: tingting@dcs.bbk.ac.uk).}
\thanks{Y. Cao is with the Key Laboratory of High Confidence Software Technologies,
Peking University, Ministry of Education, Beijing 100871, China, and also with
the Institute of Software, School of Electronics Engineering and Computer Science, Peking University, Beijing 100871, China (e-mail: caoyz@pku.edu.cn).}
}

\maketitle

\begin{abstract} Behaviour distances to measure the resemblance of two states in a  (nondeterministic) fuzzy transition system have  been proposed recently in the literature. Such a distance, defined as a pseudo-ultrametric over the state space of the model, provides a quantitative analogue of bisimilarity. In this paper, we focus on the problem of computing these distances. We first extend the definition of the pseudo-ultrametric
by introducing discount such that the discounting factor being equal to 1 captures the original definition. We then provide polynomial-time algorithms to calculate the behavioural distances, in both the non-discounted and the discounted setting. The algorithm is strongly polynomial in the former case. Furthermore, we give a polynomial-time algorithm to compute bisimulation over fuzzy transition systems which captures the distance being equal to 0.
\end{abstract}

\section{Introduction}\label{sec:intro}
\PARstart{F}{uzzy} automata and fuzzy languages are standard computational devices for modelling uncertainty and imprecision due to fuzziness. Classical fuzzy automata are \emph{deterministic}, namely, when in the current state and reading a symbol, the automaton can only move to a unique next (fuzzy) state. In \cite{CaoE12}, Cao {\it et al.} argued that \emph{nondeterminism} is essential for modelling certain aspects of system, such as scheduling freedom, implementation freedom, the external environment, and incomplete information. Hence, they introduced nondeterminism into the model of fuzzy automata, giving rise to \emph{nondeterministic fuzzy automata}, or more generally, (nondeterministic) \emph{fuzzy transition systems}. 

In general, system theory mainly concerns modelling systems and analysis of their properties. One of the fundamental questions studied in system theory is regarding the notion of equivalence, i.e., when can two systems be deemed the same and when
can they be inter-substituted for each other? In the classical investigation in concurrency theory, \emph{bisimulation}, introduced
by Park and Milner \cite{Mil89}, is a ubiquitous notion of equivalence which has become one of the primary tools in the ana\-ly\-sis
of systems: when two systems are bisimilar, known properties are readily transferred from one system to the other.
However, it is now widely recognised that traditional equivalences are not a robust concept in the presence of \emph{quantitative}
(i.e. numerical) information in the model (see, e.g., \cite{DGJP04}).
Instead, it should come up with a more robust approach to distinguish system states. To accommodate this, researchers have borrowed from pure mathematics the notion of metric. A metric is often defined as a function that associates some \emph{distance} with a pair of elements. Here, it is
exploited to provide a measure of the discrepancy between two states that are not exactly bisimilar. 
Probabilistic systems and fuzzy \newadd{transition} systems are two typical examples of systems featuring quantitative nature.
For probabilistic systems, the notion of distance in terms of pseudo-metrics has been studied extensively (cf.\ the related work).
For fuzzy \newadd{transition} systems, Cao \emph{et al} \cite{CaoSWC13} proposed a similar notion 
which serves as an analogue of those in probabilistic systems. Technically, a \emph{pseudo-ultrametric}, instead of a pseudo-metric, was adopted.
We refer the readers to Section~\ref{sec:prel} for formal definitions.

Having a proper definition of distance at hand, the next natural question is: how to compute it for a given pair of states? This
raises some algorithmic challenges. For probabilistic systems, different algorithms have been provided for a variety of stochastic models (cf the \emph{related work}). However, to the best of our knowledge, little is known as for the corresponding algorithms in fuzzy \newadd{transition} systems. Indeed, in \cite{CaoSWC13} 
this was left as an open problem, which is the main focus of the current paper. 

\newadd{On a different matter, \emph{discounting} (or inflation) is a
fundamental notion in economics and has been studied in, among others,  Markov decision processes as well as
game theory. Discounting represents the difference in importance between the future values and the present values. For instance, assuming a real-valued discount factor $0 < \gamma < 1$. A unit payoff is 1 if
the payoff occurs today, but it becomes $\gamma$ if it occurs tomorrow, $\gamma^
2$
if it occurs the day after tomorrow,
and so on.  When $\gamma=1$, the value is \emph{not} discounted. Discounting has a natural place in system engineering; as a simple example, a potential bug in the far-away future is less troubling than a potential bug today \cite{AlfaroHM03}. In other words,  discounting models preference for shorter solutions. 

We introduce discounting into the distance definition for fuzzy \newadd{transition} systems, as done in probabilistic systems \cite{BreugelSW08}. This is complementary to the definition given in \cite{CaoSWC13}. In a nutshell, when measuring the distance between two states, the distance of their one-step successors are $a$ times less important, and the distance between their two-step successors are $a^2$ times less important, etc. 
 
}
\medskip

\noindent\textbf{Contributions.} The main contributions of this paper are as follows:

\begin{itemize}
\item[(1)] We extend the pseudo-ultrametric definition
given in \cite{CaoSWC13} for non-discounted setting to the \emph{discounted} setting;

\item[(2)] We present polynomial-time algorithms to compute the behavioural distance, in both non-discounted (i.e., the original definition in \cite{CaoSWC13}) and discounted setting (defined in the current paper).

\item[(3)] We give a polynomial-time algorithm to compute the bisimulation defined in \cite{CaoCK11}.
\end{itemize}

Some explanations are in order.
Regarding (1), remark that the definition in \cite{CaoSWC13} is given in the \emph{non-discounted} setting, where the present distances and the distances in future are equally weighted. In our setting, the discounting will be taken into consideration. 
Regarding (2), the basic ingredient of our algorithms is the standard ``value iteration" procedure \emph{\`{a} la} Kleene (Kleene fixpoint theorem).
To qualify a polynomial-time algorithm, we show two facts: (i)  For each iteration, it only needs polynomial time. Note that according to the definition of pseudo-ultrametric, each step requires to solve a (non-standard) mathematical programming problem (cf.\ Section~\ref{sec:prel}). We show this can be done in polynomial-time. This part is identical for both discounted and non-discounted cases. (ii)  The number of iterations is polynomially bounded. In the non-discounted case, this is done by inspecting the possible values appearing in each iteration. For the discounted case, unfortunately this does not hold. Instead, our strategy is to firstly compute an approximation of the sought value, and then apply the continued fraction algorithm to obtain the precise value. To the best of our knowledge, we are not aware of any previous work on polynomial algorithms for computing behaviour distances in fuzzy \newadd{transition} systems.

Fuzzy \newadd{transition} systems are known as \emph{possibility systems} which are closely related to the \emph{probabilistic} systems. Our algorithm and its analysis reveal some interesting difference between these two types of models,  especially in the non-discounted case. 
Indeed, the scheme used in the paper cannot yield a polynomial-time algorithm for discrete-time Markov chains: There is an explicit example showing that it might take exponentially many iterations to reach the fixpoint; see \cite{ChenBW12}. As a matter of fact, for discrete-time Markov chains (which are the counterpart of deterministic fuzzy \newadd{transition} systems), polynomial-time algorithms do exist, but one has to appeal to linear programming \cite{ChenBW12}. This, however, does not provide a strongly polynomial-time algorithm\footnote{It is a long-standing open problem whether linear programming admits a strongly polynomial-time algorithm.}.  
Even worse, for Markov decision processes (which are the counterpart of nondeterministic fuzzy \newadd{transition} systems), the best known upper-bound is $\mathrm{NP}\cap \mathrm{co}$-$\mathrm{NP}$ \cite{Fu12}\footnote{A (weakly) polynomial-time algorithm in this case would resolve a long-standing open problem on simple stochastic games for almost 30 years.}. In contrast, here we give a strongly polynomial-time algorithm for (nondeterministic) fuzzy \newadd{transition} systems.

Regarding (3), note that two states are bisimilar if and only if the distance between them is 0 (regardless of discounted or undiscounted cases). Hence one can use the algorithms in (2) to check whether two states are bisimilar or not. However, the algorithm we present is much more efficient, and is symbolic in the sense that very little numerical information is needed. It is an adaptation of the partition-refinement algorithm for Kripke structures (or labelled transition systems; see e.g. \cite{BK08}). The bisimulation checking algorithm can be regarded as a preprocessing for the algorithm to compute the distance, as one can first identify those pair of states whose distance is 0, which might accelerate the process considerably.

\medskip

\noindent\textbf{Related work.}

\newadd{
\paragraph{Fuzzy systems, fuzzy automata and fuzzy transition systems} Conventionally, fuzzy systems are mainly referred to as fuzzy rule based systems where fuzzy states (outputs) evolve over time under some (maybe fuzzy) controls. In this paper, we are mainly interested in a type of fuzzy system models which are based on fuzzy automata \cite{Wee67}.  
Typically, fuzzy automata are considered to be acceptors of fuzzy languages. However, for the purpose of the current paper, we consider fuzzy transition systems, which are, in a nutshell, nondeterministic fuzzy automata without accepting states. Hence, we disregard the language aspect of fuzzy automata, but focus on their dynamics. 
}
\paragraph{Metrics on other types of systems}
Giacalone \emph{et al.}~\cite{GJS90} were the
first to suggest a metric between \emph{probabilistic transition systems}
to formalise the notion of distance between processes.
Subsequently, \cite{DGJP04} studied a logical pseudometric for
\emph{labelled Markov chains}. 
A similar pseudometric was defined in 
\cite{BW05} via the terminal coalgebra of a functor based on a
metric on the \emph{space of Borel probability measures}.
\cite{DJGP02} 
dealt with \emph{labelled
concurrent Markov chains}. 
\cite{DCPP06} considered a slightly more
general framework, called \emph{action-labeled quantitative
transition systems}. They defined a pseudometric which was an
adaptation of the one in \cite{DJGP02}.
Furthermore \cite{FernsPP11} considered pseudometric over \emph{Markov decision processes} with a continuous state space. 

\paragraph{Algorithms for calculating metrics}
Apart from the work discussed above, \cite{BreugelW01} gave an approximation algorithm based on linear programming and iteration. 
\cite{BreugelSW08} proposed an algorithm for Markov chains, based on the first-order theory of reals, which was extended to simple probabilistic automata in \cite{ChenHL09}. These algorithms are not optimal. \cite{TDZ11} also presented an algorithm for computing distance between probabilistic automata. However, their definition was considerably different from what is widely adopted in literature.

\paragraph{Equivalence and Metrics in fuzzy systems}Relate to the fuzzy transition systems, 
different notions of bisimulation and simulation have been introduced into traditional fuzzy automata \cite{CiricIDB12,janvcic2014weak}, weighted automata \cite{Buchholz08}, and quantitative transition systems \cite{pan2015lattice}. \'{C}iri\'{c} \emph{et al.} \cite{CiricIJD12} proposed algorithms to compute these relations. Recently, Deng and Wu \cite{deng2014modal} provided a modal cha\-racterisations of fuzzy bisimulation. As an application of fuzzy bisimulation theory, Deng and Qiu \cite{deng2015supervisory} developed the supervisory control of fuzzy discrete-event systems based on simulation equivalence. \newadd{
To the best of our knowledge, this is the \emph{first} paper to study (polynomial) algorithms of computing the distances between two fuzzy transition systems. }

%

\medskip

\noindent\textbf{Structure of the paper.} This paper is set up as follows. In Section~\ref{sec:prel}, we present some background knowledge.
In Section\,\ref{sec:algo}, \ref{sec:undis} and \ref{sec:dis} we
provide two polynomial-time algorithms for the non-discounted and discounted case, respectively. The correctness of the algorithms is also shown.
In Section~\ref{sec:bisi} we propose a polynomial time algorithm to compute the bisimulation. We conclude our work in Section~\ref{sec:conc}.

\section{Preliminaries} \label{sec:prel}

We write $\mb{Q}$ for the set of rationals. Let $X$ be a finite set. A \emph{fuzzy subset} (or simply \emph{fuzzy set}) of $X$ is a function $\mu:X\rightarrow [0,1]$. Such functions are called \emph{membership functions}; intuitively the value $\mu(x)$ captures the degree of membership of $x$ in $\mu$. A fuzzy (sub)set of $X$ can be used to formally represent a \emph{possibility distribution} over $X$. 

The \emph{support} of a fuzzy set $\mu$ is defined as $\supp(\mu)=\{x\mid \mu(x)>0\}$. If $\supp(\mu)$ is finite, we adopt the Zadah's notation. Namely, assuming $\supp(\mu)=\{x_1,x_2,...,x_n\}$, we write $\mu$ as:
\[\mu=\frac{\mu(x_1)}{x_1}+\frac{\mu(x_2)}{ x_2}+\cdots \frac{\mu(x_n)}{x_n}.\]

We write $\mc{F}(X)$ and $\Power(X)$ for the set of all fuzzy subsets and the power set of $X$ respectively.  For any $\mu, \eta\in \mc{F}(X)$, we say that $\mu$ is contained in $\eta$ (or $\eta$ contains $\mu$), denoted by $\mu\subseteq \eta$, if $\mu(x)\leq\eta(x)$ for all $x\in X$. Note that $\mu=\eta$ if both $\mu\subseteq \eta$ and $\eta\subseteq\mu$. A fuzzy set $\mu$ of $X$ is \emph{empty} if $\mu(x)=0$ for any $x\in X$. We usually write $\emptyset$ for the empty fuzzy set.


For any family $\{\lambda_i\}_{i\in I}$ of elements in $[0,1]$, we write $\bigvee_{i\in I} \lambda_i$ or $\vee\{\lambda_i\mid i\in I\}$ for the supremum of $\{\lambda_i\mid i\in I\}$, and correspondingly $\bigwedge_{i\in I} \lambda_i$ or $\wedge\{\lambda_i\mid i\in I\}$ for the infimum. Note that if $I$ is finite, $\bigvee_{i\in I} \lambda_i$ and $\bigwedge_{i\in I} \lambda_i$ are the greatest element and the least element of $\{\lambda_i\mid i\in I\}$, respectively. 
For any $\mu\in \mF(X)$ and $U\subseteq X$, $\mu(U)$ stands for $\bigvee_{x\in U} \mu(x)$.

\subsection{Fuzzy Transition Systems}

\begin{definition}[\cite{CaoSWC13}]
A \emph{fuzzy transition system} (FTS) is a tuple
$\mathcal{M}=(S, A, \delta)$ where
\begin{itemize}
\item
$S$ is a finite set of \emph{states},
\item
$A$ is a finite set of \emph{labels},
\item
$\delta:S\times A \rightarrow \mc{P}(\mc{F}(S))$ is a \emph{fuzzy transition function}.
\end{itemize}
\end{definition}

Given an FTS $(S,A,\delta)$ and $s\in S$, $a\in A$, we say $s\mv{a}\mu$ is a \emph{fuzzy transition} if $\mu\in \delta(s,a)$.
An FTS is \emph{finite} if both $S$ and $A$ are finite. Throughout this paper, we only consider finite FTSs. Let $\Act(s) = \{a\in A\,|\, \exists \mu\in \mF(S).\ s\mv{a}\mu\}$ be the set of actions enabled in state $s$.

\begin{figure}[t]
\begin{center}\scalebox{0.65}{
\includegraphics[scale=0.9]{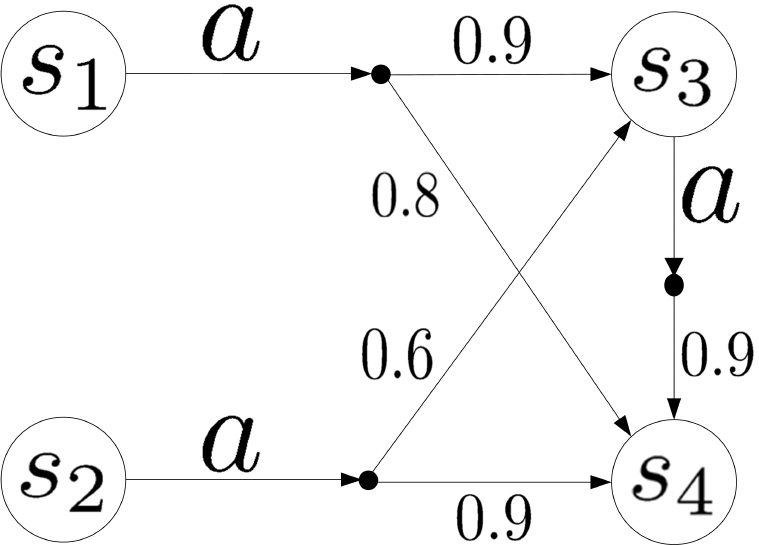}
}
\caption{Fuzzy transition system $\mM$}\label{fig:fts}\end{center}
\end{figure}

\newadd{Below we leverage the example, originally given in \cite{CaoSWC13}, to illustrate the FTS.
\begin{example}
The FTS $\mM$ is depicted in Fig.\,\ref{fig:fts}, where $S=\{s_1,s_2,s_3,s_4\}$, $A=\{a\}$ and the transitions are $s_1\mv{a}\mu$, $s_2\mv{a}\eta$ and $s_3\mv{a}\nu$. Note that here $\mu=\frac{0.9}{s_3}+\frac{0.8}{s_4}$, $\eta=\frac{0.6}{s_3}+\frac{0.9}{s_4}$, and $\nu=\frac{0.9}{s_4}$. \hfill $\square$
\end{example}
}

For complexity consideration, we need to measure the size of $\mM$. Different computational models have different size measures. For the arithmetic model of computation (for instance, \emph{unit-cost arithmetic Random Access Machine}) we define $|\mathcal{M}|$ to be   the sum of the size of the state space and the number of the entries of the transition functions. Formally $|\mathcal{M}|=|S|+\sum_{s\in S, a\in A, \mu\in \delta(s,a)}|\supp(\mu)|$, where $|\cdot|$ denotes the cardinality of a set. For the Turing machine model, we define  $||\mM||$  taking into account  the number of bits to encode the rational numbers in $\mM$. Formally $||\mM||= |S|+\sum_{s\in S, a\in A, \mu\in \delta(s,a)}||\supp(\mu)||$.  Here for each finite set $X\subseteq \mb{Q}$, $||X||=\sum_{x\in X} ||x||$, where $||x|| $ is the number of bits to encode $x$ in binary (rational numbers are represented as a fraction). Obviously $|\mM|\leq ||\mM||$.

\newadd{
\begin{example}
For the FTS $\mM$ in Fig.\,\ref{fig:fts}, $|\mM|=4+2+2+1=9$, as $|S|=4$, $|\supp(\mu)|=|\supp(\eta)|=2$, and $|\supp(\mu)|=1$. Whereas $||\supp(\mu)||=||\frac{9}{10}||+||\frac{8}{10}||=\lceil \log 9\rceil +\lceil \log 10\rceil+\lceil 
\log 8\rceil+\lceil \log 10\rceil=15$. Note that here $\log$ is to base $2$ and $\lceil \log 9\rceil +\lceil \log 10\rceil$ bits are needed to encode $0.9$ in binary. \hfill $\square$
\end{example}
}

\subsection{Behavioural Metrics}
\begin{definition} Let $X$ be a nonempty set. A function $d:X\times X\to [0,1]$ is a \emph{pseudo-ultrametric} on $X$ if for all $x,y,z\in X$:
\begin{enumerate}
\item $d(x,x)=0$;
\item $d(x,y)=d(y,x)$; and
\item $d(x,z)\leq d(x,y)\vee d(y,z)$.
\end{enumerate}
\end{definition}
The pair $(X,d)$ is a pseudo-ultrametric space. For simplicity we often write $X$ instead of $(X, d)$. 
%
In the paper we only consider $[0,1]$-valued pseudo-ultrametrics. This, however, is without loss of generality; see \cite{CaoCK11} for discussions.

Let $\mD(S)$ be the set of all pseudo-ultrametrics on $S$. For any $d\in \mD(S)$, we lift it to a pseudo-ultrametric on $\mF(S)$, as follows.
\begin{definition}[Lifting]\label{def:lifting}
Let $d\in \mD(S)$. For any $\mu,\eta\in\mF(S)$, if $\mu(S)\neq \eta(S)$, we define $\hat{d}(\mu,\eta)=1$; otherwise, we define $\hat{d}(\mu,\eta)$ as the value of the following mathematical programming problem (MP):
\begin{align}
&\textrm{minimise} & &\bigvee_{s,t\in S}(d(s,t)\wedge x_{st})\\
&\textrm{subject to} &&\bigvee_{t\in S} x_{st}=\mu(s)\quad \forall s\in S\\
                                & &&\bigvee_{s\in S}x_{st}=\eta(t)\quad \forall t\in S\\
                                & && x_{st}\geq 0\quad \forall s,t\in S
\end{align}
\end{definition}

It is shown in \cite[Theorem~1]{CaoCK11} that for each $d\in \mD(S)$, $\hat{d}$ is a pseudo-ultrametric on $\mF(S)$.

\begin{definition} \label{def:order}
We define the order $\preceq$ on $\mD(S)$ as
\[d_1\preceq d_2 \textrm{ if } d_1(s,t)\leq d_2(s,t) \textrm{ for all } s,t\in S.\]
\end{definition}
A partially ordered set $(X,\leq)$ is a \emph{complete lattice} if every subset of $X$ has a supremum and an infimum in $(X,\leq)$. It can be easily shown  that $(\mD(S),\preceq)$ is a complete lattice, following the same argument in \cite[Lemma 2]{CaoCK11}.

\newadd{
\begin{example}\label{ex:distance}
Given the FTS in Fig.\,\ref{fig:fts}, let a distance measure $d_1$ on states be defined such that $d_1(s_i,s_i)=0$ and $d_1(s_i,s_j)=1$ for all $i,j\in \{1,2,3,4\}$ and $i\neq j$. Let $d_2$ be another distance measure where $d_2(s_i,s_i)=0$ and $d_2(s_i,s_j)=0.98$ for all $i,j\in \{1,2,3,4\}$ and $i\neq j$. It is easy to see that $d_2\preceq d_1$. 

We may lift $d_1$ to distributions, i.e., $\hat{d}_1(\mu,\eta)$, or $\hat{d}_1(\mu,\nu)$, etc. We will show how to compute this measure in Algorithm 1. \hfill $\square$
\end{example}
}

Now we extend $\hat{d}$ to a distance measure on the \emph{sets} of possibility distributions by appealing to the Hausdorff distance which measures how far two subsets of a metric space are from each other. Informally, the Hausdorff distance is the longest distance of either set to the
nearest point in the other set. We consider the Hausdorff distance for a pseudo-ultrametric space.

\begin{definition}\label{def:d_x_Y}
Let $(X,d)$ be a pseudo-ultrametric space. For any $x\in X$ and $Y\subseteq X$, define
\[d(x,Y)= \begin{cases}
\bigwedge_{y\in Y} d(x,y),&\textrm{if }Y\neq \emptyset\\
1,& \textrm{otherwise}
\end{cases}
\]
\end{definition}

Further, given a pair $Y,Z\subseteq X$, the \emph{Hausdorff distance} induced by $d$ is defined as
\begin{align*}
H_d(Y,Z) =
\begin{cases}
0,&\textrm{if } {Y=Z=\emptyset}\\
\big(\bigvee_{y\in Y}d(y,Z)\big)\vee\big(\bigvee_{z\in Z}d(z,Y)\big),&\textrm{otherwise}
\end{cases}
\end{align*}
 It is shown in \cite[Lemma 3]{CaoCK11} that, if $d$ is a pseudo-ultrametric on $X$,  $H_d$ is a pseudo-ultrametric on $\mP(X)$.

\newadd{
\begin{example}
Given Fig.\,\ref{fig:fts}, let $Y=\{s_1,s_2\}\subseteq S$ and $Z=\{s_3,s_4\}\subseteq S$. The distance $d_3$ is defined as follows: $d_3(s_1,s_3)=0.92$, $d_3(s_1,s_4)=0.83$, $d_3(s_2,s_3)=0.66$, $d_3(s_2,s_4)=0.75$. As a result, $d_3(s_1,Z)= \min \{d_3(s_1,s_3),d_3(s_1,s_4)\}=\min\{0.92,0.83\}=\underline{0.83}$, and $d_3(s_2,Z)= \min\{d_3(s_2,s_3),d_3(s_2,s_4)\}=\min\{0.66,0.75\}=\underline{0.66}$.
Meanwhile, $d_3(s_3,Y)=\inf\{d_3(s_3,s_1),d_3(s_3,s_2)\}=\min\{0.92,0.66\}=\underline{0.66}$, and $d_3(s_4,Y)=\min\{d_3(s_4,s_1),d_3(s_4,s_2)\}=\min\{0.83,0.75\}=\underline{0.75}$.
The Hausdorff distance induced by $d_3$ is defined as \[H_{d_3}(Y,Z)=\max\{d_3(s_1,Z),d_3(s_2,Z),d_3(s_3,Y),d_3(s_4,Y)\}=\max\{\underline{0.83},\underline{0.66},\underline{0.66},\underline{0.75}\}=0.83.\]
\hfill$\square$
\end{example}
}

Note that any $d\in \mD(S)$ induces a pseudo-ultrametric $\hat{d}$ on $\mF(S)$ which, in turn, yields a pseudo-ultrametric $H_{d}$ on $\mP(\mF(S))$. We are now in a position to define a functional $\Delta$ on $\mD(S)$.
\begin{definition}\label{def:Delta_d_s_t}
The functional $\Delta:\mD(S)\to\mD(S)$ is defined as follows. For any $d\in\mD(S)$, $\Delta(d)$ is given by
\[\Delta(d)(s,t)=\gamma\cdot \bigvee_{a\in A}H_{{d}}\big(\delta(s,a),\delta(t,a)\big)\]
for all $s,t\in S$, where $\gamma\in (0,1]$ is the \emph{discounting factor}.
\end{definition}

In case $\gamma=1$, it is shown in \cite[Lemma 4]{CaoCK11} that the functional $\Delta:\mD(S)\to\mD(S)$ is monotonic with respect to the partial order $\preceq$.  One can easily see that this holds as well in the case that $\gamma\in (0,1)$.
%
%
By the Tarski's fixpoint theorem, $\Delta$ admits a \emph{least} fixpoint $\Delta_{\min}$ given by
\[\Delta_{\min}=\sqcap\,\{d\in\mD(S)\mid \Delta(d)\preceq d\}.\]
This is what we are after as a distance measure on the pairs of states in an FTS.

\begin{definition} \label{def:distance}
Let $(S,A,\delta)$ be an FTS and $\gamma\in(0,1]$ be the discounting factor. For any $s,t\in S$, the \emph{behavioural distance} between $s$ and $t$, denoted $d^\gamma_f(s,t)$ ($f$ stands for fixpoint), is defined as \[d_f^\gamma(s,t)=\Delta_{\min}(s,t).\]
\end{definition}

\begin{remark}
  In \cite{CaoCK11}, the order $\preceq$ is defined in the reverse direction as in Definition~\ref{def:order}, i.e., $d_1\preceq d_2$ if  $d_1(s,t)\geq d_2(s,t)$. Accordingly, $d_f^\gamma$ is defined as the greatest fixpoint (as opposed to the least fixpoint here). This is used to mimic the bisimulation which is commonly defined as the greatest fixpoint in the literature. Nevertheless these two definitions are equivalent.
\end{remark}

\newadd{
\begin{remark}[Discounting]
The transitions could be written as $s\mv{a}\mu$ and $t\mv{a}\eta$. In other words, the one-step successors of $s$ and $t$ are the distributions $\mu$ and $\eta$. The distance between $s$ and $t$ is calculated by the Hausdorff distance between their one-step successors. As it is one step away from $s$ and $t$, the contribution of the distance will be discounted by $\gamma$.  
\end{remark}
}

\section{Polynomial Algorithms for Computing Distances} \label{sec:algo}

In this section, we provide an algorithm sketch to compute the behaviour distance defined in Definition\,\ref{def:distance} for the non-discounted case ($\gamma=1$) and discounted case ($0<\gamma<1$). The basic ingredient of these algorithm is the standard iteration \emph{\`{a} la} Kleene, as they are both defined as the least fixpoint of a monotonic function.

\begin{proposition} \label{prop:kleene}
Let $(S,A,\delta)$ be an FTS and $\gamma$ be the discounting factor. Define $\Delta^0(\bot)=\bot$ and $\Delta^{n+1}(\bot)=\Delta(\Delta^n(\bot))$, where $\bot$ is given by $\bot(s,t)=0$ for all $s,t\in S$. Then $d^\gamma_f=\Delta_{\min}=\lim_{n\rightarrow \infty}\Delta^n(\bot)= \sqcup\{\Delta^n(\bot)\mid n\in \Nats\}$.
\end{proposition}

As mentioned in Section~\ref{sec:intro}, Proposition~\ref{prop:kleene} does \emph{not} yield an outright polynomial-time algorithm. 
For our purpose, essentially one has to show that 
\begin{itemize}\item[(1)] 
the (non-standard) mathematical programming (MP) problem can be solved in polynomial time; and 
\item[(2)] it only requires polynomially many iterations to reach the fixpoint. 
\end{itemize}
In the sequel, we will show both are indeed the case. For (1), we shall present a polynomial-time algorithm that, given $d$ and $s,t\in S$, compute
\[\Delta(d)(s,t)=\gamma\cdot\bigvee_{a\in A}H_{{d}}(\delta(s,a),\delta(t,a)).\] 
Intuitively, $\Delta(d)(s,t)$ is the new distance of $s$ and $t$ after applying the functional $\Delta$ to function $d$ (one iteration). This suffices to show that each iteration $\Delta$ can be done in polynomial time in either case. For (2)  it turns out that the non-discounted and discounted case demand different arguments which we will provide in Section~\ref{sec:undis} and Section~\ref{sec:dis}, respectively.

\section{The non-discounted case}\label{sec:undis}
In this section, we consider the non-discounted case, i.e., $\gamma=1$.
\subsection{The construction of an MP problem} \label{sec:MP}
We start with some simple observations. Given any $\mc{M}$,  we write $\Theta_\mc{M}$ to be
\[\{\mu(s)\,|\, {s\in S,} \ {\mu\in\mF(S)}\}\cup\{0,1\}.\]
Clearly $|\Theta_{\mM}|\leq |\mM|$. Intuitively, $\Theta_\mM$ is the set of values appearing in $\mM$ as degrees of membership as well as $0$ and $1$. 

\newadd{
\begin{example}
Given the FTS $\mM$ in Fig.\,\ref{fig:fts}, $\Theta_\mM=\{0,0.6,0.8,0.9,1\}$. \hfill $\square$
\end{example}
}

The following lemma has been implied in \cite[p.\,738, Remark 1]{CaoSWC13} and is rather straightforward. Hence the proof is omitted.
\begin{lemma} \label{lem:valuerange}
Let $\Delta(d)(s,t){=}\bigvee_{a\in A}H_{{d}}(\delta(s,a),\delta(t,a))$, for any $s,t\in S$. It holds that $\Delta(d)(s,t)\in \Theta_{\mM}\cup \{d(s,t)\mid s,t\in S \}$.
\end{lemma}
%

\newadd{We note that this observation does not hold in the discounted case.} Owning to this observation, to calculate $\Delta(d)(s,t)$, we only need to test whether each member of $\Theta_{\mM}\cup \{d(s,t)\mid s,t\in S \}$ can be attained and the least attained value is the sought one. Hence we can reduce the optimisation problem to the
feasibility testing of a system of equations $\Pi$ of the following form:
\[\begin{cases}
\bigvee_{1\leq i,j\leq n} (d_{ij}\wedge x_{ij}) = c \qquad (*)\\
x_{11}\vee x_{12}\cdots\vee x_{1k_1}=a_1\\
\qquad\qquad\vdots\\
x_{m1}\vee x_{m2}\cdots\vee x_{mk_m}=a_m
\end{cases},\]
where $c\in\Theta_\mM\cup\{d(s,t)\mid s,t\in S \}$. To this end,  we first rewrite Eq.\,$(*)$ as \[\vee Y=c, \mbox { where }Y=\{x_{ij}\mid d_{ij}\geq c\}.\qquad(**)\label{eq:star} \]
%
and thus obtain $\Pi'$ as
\begin{equation} \label{eq:piprime}
\begin{cases}
\vee Y=c, \mbox { where }Y=\{x_{ij}\mid d_{ij}\geq c\}\\
x_{11}\vee x_{12}\cdots\vee x_{1k_1}=a_1\\
\qquad\qquad\vdots\\
x_{m1}\vee x_{m2}\cdots\vee x_{mk_m}=a_m
\end{cases}\end{equation}

\newadd{Below we leverage the example FTS in Fig.\,\ref{fig:fts} to illustrate the construction.
\begin{example}\label{ex:equations}
Recall that here $\mu=\frac{0.9}{s_3}+\frac{0.8}{s_4}$ and $\eta=\frac{0.6}{s_3}+\frac{0.9}{s_4}$.
In this example, we show given $d_1$ such that $d_1(i,i)=0$ for all $i\in\{1,..,4\}$ and $d_1(i,j)=1$ for all $i\neq j$ and $i,j\in\{1,...,4\}$. Here, $\Theta_\mM\cup \{d(s,t)\mid s,t\in S \}=\{0,0.6,0.8,0.9,1\}$. In this example, we pick $c=0.6$, and the system $\Pi'$ of equations is as follows:

\begin{equation}\label{eq:example_eq}
\begin{cases}
x_{11}\vee x_{22}\vee x_{33}\vee x_{44}=0.6 \\
x_{11}\vee x_{12}\vee x_{13}\vee x_{14}=0 & \qquad\mu(s_1)=0\\
x_{21} \vee x_{22}\vee x_{23}\vee x_{24}=0& \qquad\mu(s_2)=0\\
x_{31} \vee x_{32}\vee x_{33}\vee x_{34}=0.9& \qquad\mu(s_3)=0\\
x_{41} \vee x_{42}\vee x_{43}\vee x_{44}=0.8& \qquad\mu(s_4)=0\\
x_{11} \vee x_{21}\vee x_{31}\vee x_{41}=0& \qquad\eta(s_1)=0\\
x_{12} \vee x_{22}\vee x_{32}\vee x_{42}=0& \qquad\eta(s_2)=0\\
x_{13} \vee x_{23}\vee x_{33}\vee x_{43}=0.6& \qquad\eta(s_3)=0\\
x_{14} \vee x_{24}\vee x_{34}\vee x_{44}=0.9& \qquad\eta(s_4)=0
\end{cases}
\end{equation}
Note that systems of equations for other values of $c$ can be constructed in the same way. \hfill $\square$

\end{example}
}

\subsection{Checking feasibility}\label{sec:fea}

\newadd{Recall that now we have a system of equations $\Pi$ (which can be rewritten as $\Pi'$). Our task is to check whether both $\Pi$ and $\Pi'$ are feasible, i.e., whether there exists a solution to make $\Pi$ and $\Pi'$ hold.}

\begin{lemma}\label{lem:correctness}
$\Pi$ is feasible \emph{iff} $\Pi'$ is feasible. 
\end{lemma}
\newadd{
\begin{proof}
It is to show that $(*)$ and $(**)$ are equivalent. We rewrite $(*)$ as $\sup_{1\leq i,j\leq n}\{(d_{ij}\wedge x_{ij})\} =c$. For all $1\leq i,j\leq n$, if $d_{ij}<c$, then $d_{ij}\wedge x_{ij}<c$. As a result, this $(d_{ij}\wedge x_{ij})$ is not contributing to $\sup_{1\leq i,j\leq n}\{(d_{ij}\wedge x_{ij})\} =c$ and can be removed. 
\end{proof}
}

Given Lemma \ref{lem:correctness}, it remains to show how to check the feasibility of a system of equations $\Pi'$ of the form
\[
\begin{cases}
x_{11}\vee x_{12}\cdots\vee x_{1k_1}=a_1\\
\qquad\qquad\vdots\\
x_{m1}\vee x_{m2}\cdots\vee x_{mk_m}=a_m
\end{cases}
\]
For this purpose, we construct a valuation for $x_i$ ($1\leq i\leq n$) as
\[x_i=\min \{a_\ell\mid x_{\ell h}=x_i \mbox{ for some } 1\leq h\leq k_{\ell}\}\]
The following lemma plays a vital role. In other words, in order to check the feasibility of $\Pi'$, it suffices to check whether $\vec{x}=(x_i)_{1\leq i\leq n}$ is a solution of $\Pi'$. This can be done in linear time.

\begin{lemma} \label{lem:correctness2}
$\Pi'$ is feasible iff $\vec{x}=(x_i)_{1\leq i\leq n}$ is a solution of $\Pi'$.
\end{lemma}

\begin{proof}
The ``if" part is obvious and we will focus on the ``only if" part. Fix any $1\leq i\leq m$.
By the definition of $\vec{x}$, we have that, for each $1\leq j\leq k_i$, $x_{ij}=\min \{a_\ell\mid x_{\ell h}=x_{ij} \mbox{ for some } 1\leq h\leq k_{\ell}\}$. Clearly $a_i\in \{a_\ell\mid x_{\ell h}=x_{ij} \mbox{ for some } 1\leq h\leq k_{\ell}\}$, and thus we have that $x_{ij}\leq a_i$. It follows that
\[x_{i1}\vee x_{i2}\cdots\vee x_{ik_i}\leq a_i.\]

Since $\Pi'$ is feasible, there must exist a solution $\vec{x}'$ of $\Pi$. For each $1\leq i\leq n$,
$x_i\leq a_\ell$ with $x_{\ell h}=x_i$ for some $1\leq h\leq k_{\ell}$. Consequently
\[\vec{x}'\leq \vec{x}.\]
It follows that for each $1\leq i\leq m$,
\[a_i = x'_{i1}\vee x'_{i2}\cdots\vee x'_{ik_1} \leq x_{i1}\vee x_{i2}\cdots\vee x_{ik_1}\leq a_i\]
Namely, $\vec{x}$ is a solution of $\Pi'$.
\end{proof}

\subsection{The algorithms}

\newadd{Based on the results in Section \ref{sec:MP} and \ref{sec:fea}, we specify three algorithms here. As illustrated in Fig.\,\ref{fig:algo}, given the current $d(s,t)$ Algorithm\,\ref{alg:hat_d_n} computes $\hat{d}(\mu,\eta)$. The results act as the input for Algorithm\,\ref{alg:d_n+1}, which calculates the updated $d(s,t)$, i.e., $\Delta(d)(s,t)$. 
In Algorithm\,\ref{alg:main}, it repeats the procedure until a fixpoint is reached.

\begin{figure}[!ht]
\includegraphics[scale=0.4]{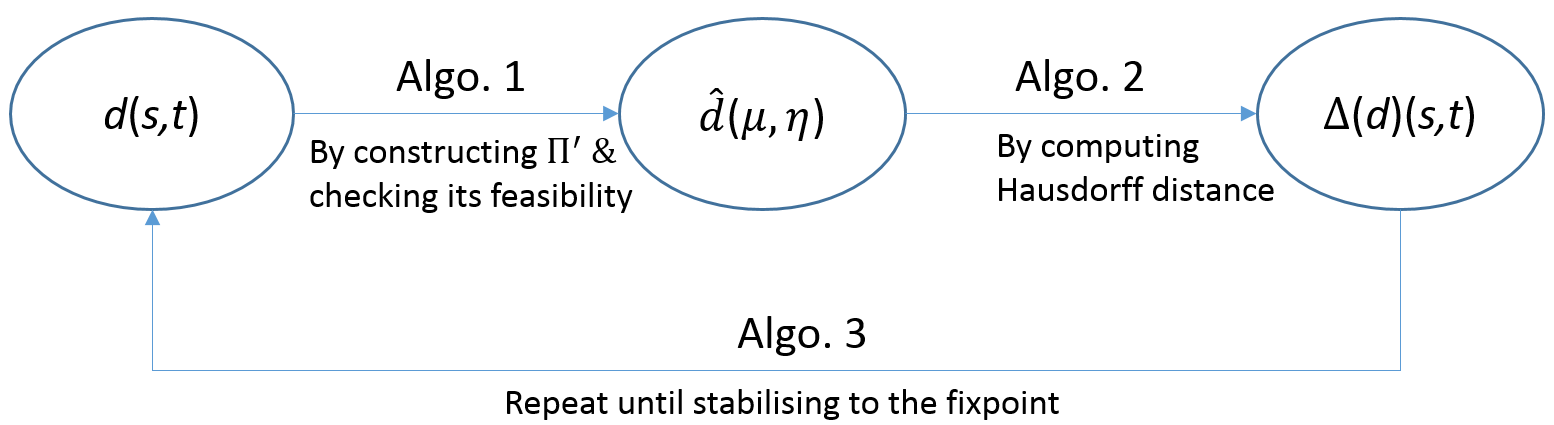}\caption{\newadd{Overview of the algoritms}}\label{fig:algo}
\end{figure}
}

\begin{algorithm}
\KwData{FTS $(S,A,\delta)$, distance $d(s,t)$ for all $s,t\in S$, and distributions $\mu,\eta\in\mF(S)$}
\KwResult{The distance $\hat{d}(\mu,\eta)$ induced by $d$}
\Begin{
Sort the set $\Theta\longleftarrow \{\mu(s),\eta(s), d(s,t)\mid s,t\in S\}\cup\{0\}$ in an \textbf{ascending} order with the $i$-th element of $\Theta$ denoted by
$\Theta_i$\;
$i\longleftarrow 1$\;
\While{$i\leq |\Theta|$}{
\tcp{Find a potential solution} 
Take the system of equations $\Pi'$ (cf.~\eqref{eq:piprime})\;
\For{\emph{\textbf{each}} pair $(s,t)$}{
$x_{st}\longleftarrow 1$\;
\For{\emph{\textbf{each}} equation $\pi$ in $\Pi'$}{
\If{$x_{st}$ appears in the left hand side of $\pi$}{
$x_{st}\longleftarrow \min\{x_{st}, \textit{ the right hand side of }\pi\}$\;
}
}
}
\tcp{Test whether the potential solution is a solution of $\Pi'$.}

%
\If{all the equations in $\Pi'$ hold}{
The potential solution is a solution of $\Pi'$\;
$\hat{d}_n(\mu,\eta)\longleftarrow\Theta_i$\;
\textbf{break}\;
}
$i\longleftarrow i+1$\;
}
}\caption{Calculate $\hat{d}(\mu,\eta)$}\label{alg:hat_d_n}
\end{algorithm}
%

%
\begin{algorithm}
\KwData{FTS $(S,A,\delta)$, 
and $s,t\in S$, distance function $d$}
\KwResult{$\Delta(d)(s,t)$}
\Begin{
\If{$Act(s)\neq  Act(t)$}{
$d\longleftarrow 1$\;
}\Else{
$d\longleftarrow 0$\;
$Act \longleftarrow Act(s)$\;
\While{$Act\neq\emptyset$}{
Take $a\in \Act$\;
\tcp{Compute Hausdorff distance}
\tcp{$H_{d}(\delta(s,a),\delta(t,a))\longleftarrow q$}
$Y\longleftarrow \delta(s,a)$\;
$Z\longleftarrow \delta(t,a)$\;
$q\longleftarrow 0$\;
\While{$Y\neq \emptyset$}{
Take $\mu\in Y$\;
$r\longleftarrow 1$\;
\While{$Z\neq \emptyset$}{
Take $\eta\in Z$\;
$r\longleftarrow r\wedge \hat{d}(\mu,\eta)$ \tcp*[l]{Call Algo.\,\ref{alg:hat_d_n}}
$Z\longleftarrow Z\setminus \{\eta\}$\;
}
$q\longleftarrow q\vee r$\;
$Y\longleftarrow Y\setminus\{\mu\}$\;
}$d\longleftarrow d \vee q$\;
$\Act\longleftarrow\Act\setminus \{a\}$\;
}
}
$\Delta(d)(s,t)\longleftarrow d$\;
}\caption{Calculate $\Delta(d)(s,t)$}\label{alg:d_n+1}
\end{algorithm}

We will continue with Example \ref{ex:equations} to show how to calculate $\hat{d_1}(\mu,\eta)$.
\begin{example}[Continued.]
According to the algorithm, $\Theta=\{0,0.6,0.8,0.9\}$. The algorithm will start from $\Theta_1=0$. It turns out this is not a feasible solution. We will move on to try $\Theta_2=0.6$. As a result, we obtain a potential solution $x_{11}=\min\{0.6,0,0\}=0$, $x_{33}=\min\{0.6,0.9,0.6\}=0.6$, $x_{34}=\min\{0.9,0.9\}=0.9$, $x_{44}=\min\{0.6,0.8,0.9\}=0.6$, $x_{43}=\min\{0.8,0.6\}=0.6$ and all the rest $x_{ij}=0$.

Now we test whether the potential solution is really a solution of $\Pi'$ in Eq.\,\eqref{eq:example_eq}. This is done by substituting the values back to Eq.\,\eqref{eq:example_eq}. It is clear that $x_{41} \vee x_{42}\vee x_{43}\vee x_{44}=0.8$, but $0\vee 0\vee 0.6\vee 0.6=0.6\neq 0.8$. So this potential solution is \emph{not} a solution to $\Pi'$.

The algorithm proceeds by checking the next candidate $\Theta_3=0.8$.
Here, we have $x_{33}=\min\{0.8,0.9,0.6\}=0.6$, $x_{34}=\min\{0.9,0.9\}=0.9$, $x_{44}=\min\{0.8,0.8,0.9\}=0.8$, $x_{43}=\min\{0.8,0.6\}=0.6$ and all the rest $x_{ij}=0$. And this potential solution \emph{is} a solution to $\Pi'$. This is to say that the distance between $\mu$ and $\eta$ induced by $d_1$ is $\hat{d_1}(\mu,\eta)=0.8$.\hfill $\square$
\end{example}

\medskip

Whereas the correctness of Algorithm\,\ref{alg:d_n+1} follows from definition and it is easy to see it only requires polynomial time. The correctness of Algorithm\,\ref{alg:hat_d_n} is far from trivial.

Lemma~\ref{lem:correctness} and Lemma~\ref{lem:correctness2} give rise to:
%
%
%
\begin{proposition} \label{prop:correctness2}
Algorithm\,\ref{alg:d_n+1} computes, for each $d$, $\Delta(d)$ in polynomial time.
\end{proposition}

We remark that this is actually a \emph{strongly} polynomial-time algorithm.
In the literature, strongly polynomial time is defined in the arithmetic model of computation. In this model, the basic arithmetic operations (addition, subtraction, multiplication, division, and comparison) take a unit time step to perform, regardless of the sizes of the operands. The algorithm runs in strongly polynomial time  \cite{GLS93} if
(1) the number of arithmetic operations  is bounded by a polynomial in $|\mM|$; and
(2) the space used by the algorithm is bounded by a polynomial in $||\mM||$.
In our case, one can easily verify (1) and (2) hold for Algorithm\,\ref{alg:d_n+1}. 




Algorithm\,\ref{alg:main} is the main procedure to compute $d^1_f(s,t)$ by an iteration of $\Delta$ starting from the least element $\bot$. The correctness of the algorithm follows from Proposition~\ref{prop:correctness2}, which also asserts that each iteration requires polynomial time only. Hence it suffices to show that only polynomial number of iterations are necessary to terminate the algorithm. 

\begin{algorithm}[!ht]
\KwData{FTS $(S,A,\delta)$}
\KwResult{The behavioural distance matrix $d_f^1$}
\Begin{
$ n\longleftarrow 0$\;
$d_0\longleftarrow \bot$\;
\Repeat{$\mathbf{D}_l= \mathbf{D}$}{
$\mathbf{D}_l\longleftarrow d_n$ \;
\For{\emph{\textbf{each}} pair $(s,t)$}{
$d_{n+1}(s,t)\longleftarrow \Delta(d_n)(s,t)$;  \tcp{Call Algo.\,\ref{alg:d_n+1}}
}
$\mathbf{D}\longleftarrow d_{n+1}$\;
$n\longleftarrow n+1$\;
}
$d_f^1=\mathbf{D}$\;
}
\caption{Calculate $d_f^1$}\label{alg:main}
\end{algorithm}

The following lemma can be obtained by induction and Lemma~\ref{lem:valuerange}.
\begin{lemma}\label{lem:Theta_beh_dist}
Given $\mM=(S,A,\delta)$, it holds that $d_f^1(s,t)\in\Theta_\mM$ for any $s,t\in S$.
\end{lemma}

The following proposition shows only polynomial number of iterations are necessary to terminate the algorithm. 

\begin{proposition}\label{prop:iter_poly}
The iteration in Algorithm\,\ref{alg:main} needs at most polynomially many steps.
\end{proposition}
\begin{proof}
We assume that $n\in \Nats$ is the smallest $n$ such that $d_n(s,t)=d_{n+1}(s,t)$. Due to Lemma\,\ref{lem:Theta_beh_dist}, $d_i(s,t)\in\Theta_\mM$, for any $i\in [0,n]$. Note that each $d_i$ is of dimension $|S|^2$, and thus the value of each entry $d_{st}$ must be in $\Theta_\mM$. Moreover, as $\Delta$ is monotonic, $d_{i+1}\leq d_i$. It follows that $n\leq |\Theta_\mM|\cdot |S|^2$, which is polynomial in the size of $\mM$. 
\end{proof}

We conclude this section by the following theorem, which can be easily shown by Proposition \ref{prop:correctness2} and \ref{prop:iter_poly}.

\begin{theorem} \label{thm:main1}
Given a fuzzy transition system $\mM$, and two states $s,t$, $d^1_f(s,t)$ can be computed in (strongly) polynomial time.
\end{theorem}

%
%
%

%

\section{The Discounted Case} \label{sec:dis}

In this section, we consider the discounted case, i.e., $\gamma<1$. First we remark that one cannot (at least not in a straightforward manner) follow the same approach as in Proposition~\ref{prop:iter_poly} to obtain a polynomial-time algorithm, simply because Lemma~\ref{lem:Theta_beh_dist} fails when $\gamma<1$. Instead, our strategy is to first come up with an
approximation algorithm, which, given any $\epsilon>0$, computes a $d$ such that $||d-d^\gamma_f||\leq \epsilon$. It turns out that such an approximation can be identified by applying at most $\lceil \log_{\gamma}\epsilon\rceil$ iterations.
In the sequel, we consider the $\infty$-norm for vectors, i.e., $||d_1-d_2||=\max_{s,t\in S} |d_1(s,t)-d_2(s,t)|$.
We first show that $\Delta$ is a contraction mapping. The following is a simple technical fact.

\begin{lemma}\label{lem:min_trick}
For any $z_1,z_2,t\in \Reals$, it holds that $\min(z_1,t)-\min(z_2,t)\leq |z_1-z_2|$.
\end{lemma}
\begin{proof}
Observe that $\min(z,t)=\dfrac{|z+t|-|z-t|}{2}$.
It then follows that
\begin{align*}
&\ \min(z_1,t)-\min(z_2,t) \\
=&\ \dfrac{|z_1+t|-|z_1-t|+|z_2-t|-|z_2+t|}{2}\\
\leq &\ \dfrac{|(z_2-t)-(z_1-t)|+|(z_1+t)-(z_2-t)|}{2}\\
=&\ |z_1-z_2|.
\end{align*}
\end{proof}

For simplicity, given $\mu$ and $\eta$, we write $U_{\mu,\eta}$ for the set of $\{x_{uv}\}_{u,v\in S}$ such that
\[
\begin{cases}
\bigvee_{v\in S} x_{uv} = \mu(u) & \forall u\in S\\
\bigvee_{u\in S} x_{uv}=\eta(v) &\forall v\in S\\
x_{uv}\geq 0 & \forall u,v\in S
\end{cases}
\]

\begin{lemma}\label{lem:Delta_d}
For any $d,d':S\times S \to [0,1]$, it holds that
\[||\Delta(d)-\Delta(d')|| \leq \gamma\cdot ||d-d'||. \]
\end{lemma}
\begin{proof}
By definition, $$||\Delta(d)-\Delta(d')|| =\max_{s,t\in S}|\Delta(d)(s,t)-\Delta(d')(s,t)|.$$ Fix two states $s$ and $t$. We have the following:
\begin{eqnarray*}
&&|\Delta(d)(s,t)-\Delta(d')(s,t)|\\
&=&|\gamma\cdot\bigvee_{a\in A}H_{d}(\delta(s,a),\delta(t,a))-\gamma\cdot\bigvee_{a\in A} H_{d'}(\delta(s,a),\delta(t,a))|\\
&\leq &\ \gamma\cdot |H_{d}(\delta(s,a^*),\delta(t,a^*))-H_{d'}(\delta(s,a^*),\delta(t,a^*))|
\ [\mbox{where } a^*=\arg\max_{a\in A} H_{d}(\delta(s,a),\delta(t,a))] \\
&\leq &\ \gamma\cdot |\hat{d}(\mu^*,\eta^*)-\hat{d'}(\mu^*,\eta^*)|\ [\mbox{where } (\mu^*,\eta^*)=\arg H_{d}(\delta(s,a^*),\delta(t,a^*))]\\ 
&= &\ \gamma\cdot |\min_{\vec{x}\in U_{\mu^*,\eta^*}}\bigvee_{u,v\in S}(d(u,v)\wedge x_{uv})-\min_{\vec{x}\in U_{\mu^*,\eta^*}}\bigvee_{u,v\in S}(d'(u,v)\wedge x_{uv})|\\
&\leq &\ \gamma\cdot |\bigvee_{u,v\in S}(d(u,v)\wedge y_{uv})-\bigvee_{u,v\in S}(d'(u,v)\wedge y_{uv})|\ [\mbox{where } \vec{y}=\arg\min_{\vec{x}\in U_{\mu^*,\eta^*}} \bigvee_{u,v\in S}(d(u,v)\wedge x_{uv})]\\
&\leq &\ \gamma\cdot |d(u^*,v^*)\wedge y_{u^*v^*}-d'(u^*,v^*)\wedge y_{u^*v^*}|\ [\mbox{where }(u^*, v^*)= \arg\max_{u,v\in S} d(u,v)\wedge y_{uv}] \\ 
&\leq &\ \gamma\cdot |d(u^*,v^*)-d'(u^*,v^*)|\ [\mbox{By Lemma~\ref{lem:min_trick}}] \\
&\leq &\ \gamma\cdot ||d-d'||
\end{eqnarray*}
Namely, for any $s,t\in S$, $$|\Delta(d)(s,t)-\Delta(d')(s,t)|\leq \gamma\cdot ||d-d'||.$$ As a result, $$||\Delta(d)-\Delta(d')||=\max_{s,t\in S}|\Delta(d)(s,t)-\Delta(d')(s,t)|\leq \gamma\cdot ||d-d'||.$$
\end{proof}

Lemma~\ref{lem:Delta_d} reveals that $\Delta$ is a contraction mapping, hence by the Banach fixpoint theorem, $d^\gamma_f$ is not only the least, but also the unique fixpoint of $\Delta$.

\begin{theorem}\label{thm:discount_n_steps}
Given any FTS $\mM$ with discounting factor $\gamma\in (0,1)$, let $N=\lceil \frac{\log \epsilon}{\log \gamma}\rceil$. Then $||\Delta^N(d_0)-d^\gamma_f||\leq \epsilon$.
\end{theorem}
\begin{proof}
First, observe that $||\Delta(d)-d^\gamma_f||\leq \gamma\cdot ||d-d^\gamma_f||$, which follows from Lemma\,\ref{lem:Delta_d} and the fact that $\Delta(d^\gamma_f)=d^\gamma_f$.

By induction, we have $$||\Delta^n(d_0)-d^\gamma_f||\leq \gamma^n\cdot ||d_0-d^\gamma_f||.$$ Hence
\[||\Delta^N(d_0)-d^\gamma_f||\leq \gamma^N\cdot ||d_0-d^\gamma_f||\leq \gamma^N \leq \epsilon.\]
\end{proof}

\begin{algorithm}[!ht]
\KwData{FTS $(S,A,\delta)$, error bound $\epsilon$, discounting factor $\gamma$}
\KwResult{The approximate behavioural distance $d_{f\epsilon}^\gamma$}
\Begin{
$ N\longleftarrow \lceil \frac{\log \epsilon}{\log r}\rceil$\;
$d_0\longleftarrow \bot$\;
$ n\longleftarrow 0$\;
\Repeat{$n>N$}{
\For{\emph{\textbf{each}} pair $(s,t)$}{
$d_{n+1}(s,t)\longleftarrow \gamma \cdot \Delta(d_n)(s,t)$; {\tcp{Call Algo.\ref{alg:d_n+1}}}
}
$\mathbf{D}\longleftarrow d_{n+1}$\;
$n\longleftarrow n+1$\;
}
$d_{f\epsilon}^\gamma=\mathbf{D}$\;
}
\caption{Calculate $d^\gamma_{f\epsilon}$}\label{alg:main2}
\end{algorithm}

Theorem~\ref{thm:discount_n_steps} states that $\Delta^N(d_0)$ approximates $d^\gamma_f$ up to $\epsilon$. It also gives a strongly polynomial \emph{approximation} algorithm up to any precision $\epsilon$, as shown in Algorithm~\ref{alg:main2}. This is almost sufficient for practical considerations.
Theoretically appealing, by the standard continued fraction algorithm \cite{GLS93} , we can
compute the \emph{exact} $d^\gamma_f$ in polynomial time as well. For this purpose, we need the following lemma:

 \begin{lemma}\label{lem:size}
 For $\gamma\in (0,1)$, $d^{\gamma}_f$ is a rational vector of size polynomial in $||\mM||$ and $||\gamma||$.
 \end{lemma}
 \begin{proof}
 For simplicity we write $d$ for $d^\gamma_f$. By definition, $d$ must satisfies
 \[d(s,t)= \gamma \cdot \hat{d}(\mu, \eta)\]
 for some $a\in A$, $\mu\in \delta(s, a)$, and $\eta\in \delta(t, a)$. Namely
 \[d(s,t)=   \gamma \cdot \bigvee_{u,v\in S}(d(u,v)\wedge x_{u,v})\]
such that
\[
\begin{cases}
\bigvee_{v\in S} x_{uv}=\mu(u) \quad \forall u\in S\\
\bigvee_{u\in S}x_{uv}=\eta(v)\quad \forall v\in S\\
x_{uv}\geq 0\quad \forall u,v\in S
\end{cases}
\]
The claim hence follows from basic linear algebra.
 \end{proof}

 \begin{theorem} \label{thm:main2}
 For a fixed $\gamma$, $d$ can be computed exactly in polynomial time in $||\mM||$.
 \end{theorem}

\begin{proof}
By Theorem~\ref{thm:discount_n_steps}, we can find $d^\gamma_f$ in polynomial time in $\mM$ and $\epsilon$ a vector that is $\epsilon$-close to $d^\gamma_f$. And by Lemma~\ref{lem:size}, $d^\gamma_f$ is a rational vector of size polynomial in $\|\mM\|$. So we can use the continued fraction algorithm \cite[Chapter~5]{GLS93} to compute
$d$ in polynomial time, as is illustrated in \cite{ChenBW12}.
\end{proof}

We remark that, unfortunately, the exact polynomial-time algorithm is \emph{not} strongly polynomial, as continued fraction algorithm is used. It is an open question whether one can obtain an exact strongly polynomial-time algorithm.


\section{Bisimulation} \label{sec:bisi}

In \cite{CaoCK11, CaoSWC13}, bisimulation over fuzzy \newadd{transition} systems was introduced.
In this section, we give a polynomial-time algorithm to decide whether two states $s,t\in S$ are bisimilar. This also yields a polynomial algorithm to check whether $d(s,t)=0$; cf \cite[p.\ 740, Theorem~4]{CaoSWC13}.

Algorithm~\ref{alg:quotient} is adapted from classical partition-refinement based algorithms for computing the bisimulation in Kriple structures or labelled transition systems. The correctness of the algorithm, as well as the analysis of efficiency, is very similar to the classical case, hence are omitted here.

The following definition of bisimulation adopts \cite[p.\ 740, Definition~9]{CaoSWC13}.

\begin{definition}
Let $\mM=(S,A,\delta)$ be an FTS. An equivalence relation $R\subseteq S\times S$ is a \emph{bisimulation} on $S$ if for any $(s,t)\in R$, $s\stackrel{a}{\rightarrow} \mu$ implies that
$t\stackrel{a}{\rightarrow} \eta$ such that $\mu(C)=\eta(C)$, for all $C\in S/R$. (Note that here $C\in S/R$ refers to the quotient set of $S$ by the equivalence relation $R$.)

States $s$ and $t$ are bisimulation equivalent (or bisimilar), denoted $s\sim t$,
if there exists a bisimulation $R$ on $\mM$ such that $(s, t) \in R$.
\end{definition}

\begin{algorithm}[!ht]
\KwData{FTS $(S,A,\delta)$}
\KwResult{Bisimulation quotient space $S/\sim$ }
\Begin{
$ \Xi\longleftarrow S$\;
$\Xi_{\textit{old}}\longleftarrow S$\;
\Repeat{$\Xi_{\textit{old}}= \Xi$}{
$\Xi_{\textit{old}}\longleftarrow \Xi$ \;
\For{\emph{\textbf{each}} $C\in \Xi_{\textit{old}} $ and $a\in A$}{
$\Xi\longleftarrow \mathit{Refine}(\Xi,C,a)$; \hfill\tcp*[h]{Call Algo.\,\ref{alg:refine}}
}
}
$S/\sim\longleftarrow\Xi$\;
}
\caption{Calculate the quotient state space of bisimulation $\sim$}\label{alg:quotient}
\end{algorithm}

\begin{algorithm}[!ht]
\KwData{FTS $(S,A,\delta)$, a state partition $\Xi$, a splitter $C\in\Xi$ and an action $a\in A$}
\KwResult{A new partition with respect to $C$ and $a$}
\Begin{
\For{\emph{\textbf{each}} $B\in \Xi $}{

$\Theta_B\longleftarrow \emptyset$\;
\For{\emph{\textbf{each}} $s\in B $}{
\If{there exists $\mu$ such that $s\mv{a}\mu$}{
\If{$\mu(C)\notin \Theta_B$}{
$\Xi\longleftarrow\Pi\setminus \{B\}\cup\{B_{\mu(C)}\}$\;
$\Theta_B\longleftarrow \{\mu(C)\}$\;
}
$B_{\mu(C)}\longleftarrow B_{\mu(C)}\cup\{s\}$\;
}
}

}
$\mathit{Refine}(\Xi,C,a)\longleftarrow \Xi$\;
}
\caption{$\mathit{Refine}(\Xi,C,a)$ -- Refine a partition $\Xi$ using a splitter $C$ and action $a$ }\label{alg:refine}
\end{algorithm}


\section{Conclusion} \label{sec:conc}
We have studied the algorithmic aspect of behavioural distance for fuzzy \newadd{transition} systems.
The pseudo-ultrametric defined in \cite{CaoSWC13} was extended to accommodate both the discounted and non-discounted settings. We then provided polynomial-time algorithms to calculate the behavioural distance in both cases. We also gave a polynomial-time algorithm to compute the bisimulation defined in \cite{CaoCK11}.

\bibliographystyle{abbrv}
\bibliography{fuzzy}
\end{document}